\documentclass[showpacs,preprintnumbers,amsmath,amssymb, superscriptaddress, twocolumn]{revtex4}%
 \usepackage{graphicx}					
\usepackage{dcolumn}					
\usepackage{bm}						
\usepackage{natbib}
\usepackage{color}
\begin{document}

\title{Spin dynamics in the frustrated three-dimensional hyperkagom{\'e} compound Gd$_{3}$Ga$_{5}$O$_{12}$ }

\author{P.~P.~Deen}
\affiliation{Institut Laue-Langevin, 6 rue Jules Horowitz, 38042 Grenoble, France}
\author{O.~A.~Petrenko}
\affiliation{1Department of Physics, University of Warwick, Coventry, CV4 7AL, United Kingdom}
\author{G.~Balakrishnan}
\affiliation{1Department of Physics, University of Warwick, Coventry, CV4 7AL, United Kingdom}
\author{B.~D.~Rainford}
\affiliation{Department of Physics and Astronomy, Southampton University, Southampton, SO17 0BJ, United Kingdom}
\author{C.~Ritter}
\affiliation{Institut Laue-Langevin, 6 rue Jules Horowitz, 38042 Grenoble, France}
\author{L.~Capogna}
\affiliation{Istituto Officina dei Materiali, IOM-CNR, OGG 6 rue J. Horowitz, 38042 Grenoble, France.}
\author{H.~Mutka}
\affiliation{Institut Laue-Langevin, 6 rue Jules Horowitz, 38042 Grenoble, France}
\author{T.~Fennell}
\affiliation{Institut Laue-Langevin, 6 rue Jules Horowitz, 38042 Grenoble, France}

\date{\today}

\begin{abstract}
We present the first neutron inelastic scattering results on the low temperature magnetic state of the three dimensional hyperkagom{\'e} compound Gd$_{3}$Ga$_{5}$O$_{12}$ (GGG). GGG is often classified as a  strongly frustrated system with a manifold of continuously connected states. 
However, in contrast to the expectation of a continuum of gap-less
excitations above a spin liquid-like ground state our results reveal three
distinct inelastic modes found at 0.04(1), 0.12(2) an 0.58(3) meV at 0.06~K. The inelastic modes can be attributed to the magnetic ground state with the lowest and highest energy excitations showing spatial dependencies indicative of dimerized short range antiferromagnetic correlations. 
Short range correlations, reminiscent of spin liquid-like order, are static within the instrumental resolution (50~$\mu$eV) and represent 82~$\%$ of the spectral weight. Longer range correlations, first observed by Petrenko {\it et al.}\cite{Petrenko1998}, develop below 0.14 ~K within the elastic cross section. The short range static correlations and dynamic components survive to high temperatures, comparable to the nearest neighbor exchange interactions.  Our results suggest that the ground state of a three dimensional hyperkagom{\'e} compound differs  distinctly from its frustrated counterparts on a pyrochlore lattice and reveals a juxtaposition of spin liquid order and strong dimerised coupling. 

\end{abstract}
 
\maketitle

In recent years it has become evident that magnetic frustration provides  an excellent path to novel and exotic magnetic order \cite{Anderson1973, Ramirez1994, Moessner2006, Gardner2010, Balents2010}.   Evocative names such as spin liquids, spin glasses and spin ice are associated with the frustration of magnetic spins.  In spin liquids the energy scale of interactions between the spins does not influence the ordering temperature and due to a manifold of degenerate states the spins remain fluctuating at temperatures much lower than the interaction energies.  An illustrative example  of a spin liquid with a large spin value S $>>$ $\frac{1}{2}$, a cooperative paramagnet, is Tb$_{2}$Ti$_{2}$O$_{7}$. Tb$_{2}$Ti$_{2}$O$_{7}$ remains disordered down to the lowest temperatures and displays a spin liquid state  \cite{Gardner1999}. The excitation spectrum reveals, in addition to crystal field excitations \cite{Gardner2001},  a continuum of fluctuating spins that slow down with decreasing temperature but remain fluctuating down to 0.05~K  \cite{Gardner2003}. A second example of a cooperative paramagnet is the kagom\'{e} antiferromagnet deuteronium jarosite which shows gapless magnetic excitations extending out to at least 20~meV with a linear temperature dependence of the spin fluctuation rate \cite{Fak2008}.  These two examples  highlight the continuum of liquid-like quasielastic scattering typically observed in a cooperative paramagnet and conform to the prediction by Moessner {\it et al.} that a  system of classical spins on some frustrated lattices will observe a linear temperature dependence of the spin fluctuation rate  \cite{Moessner1998, Moessner1998PRL}.

Contrary to expectation this work presents, to our knowledge, the first inelastic neutron scattering study on a three dimensional (3D) hyperkagom{\'e} structure,  Gd$_{3}$Ga$_{5}$O$_{12}$ (GGG) in which we show spin liquid order that is concomitant with distinct gapped modes pointing towards singlet-triplet excitations arising from short range antiferromagnetic (AF) correlations. These results thus shed new light on the spin dynamics of frustrated hyperkagom\'{e} structures.   
\begin{figure}[htp]
\begin{center}
	\includegraphics[width=6cm, angle=0.]{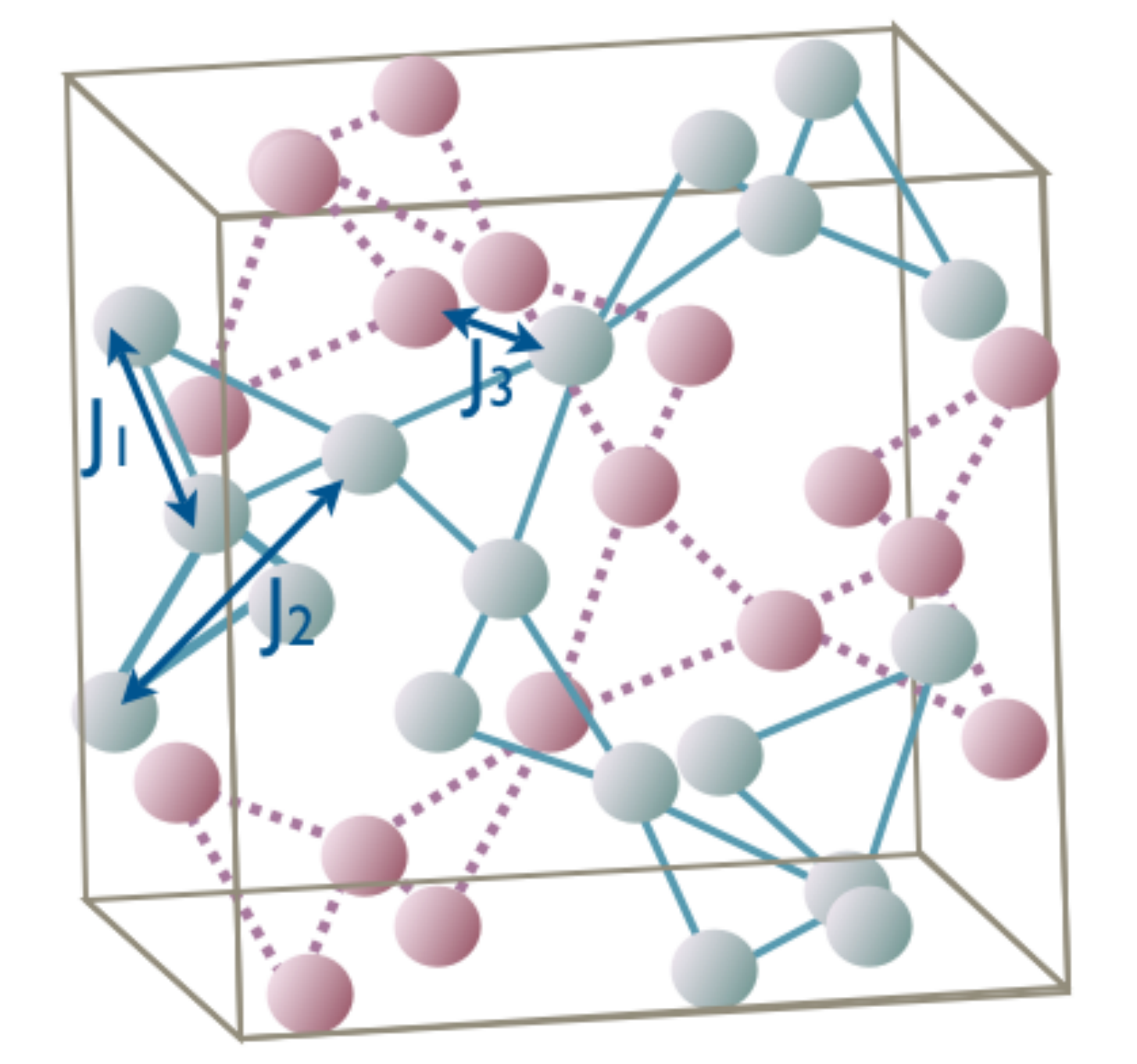}
	\caption{The garnet structure of GGG with two interpenetrating hyperkagom{\'e} lattices showing the interatomic, inter-triangular and inter hyperkagom{\'e}  exchange interactions, $J_{1}$, $J_{2}$ and $J_{3}$, respectively. For clarity only the Gd$^{3+}$ are shown.}
\label{SQW}
\end{center}
\end{figure}

Structurally, rare earth garnets such as GGG are one of very few realizations of a hyperkagom{\'e} structure, a 3D lattice of corner sharing triangles. Another recently created hyperkagom{\'e} compound is Na$_{4}$Ir$_{3}$O$_{8}$, \cite{Lawler2008}. In fact, GGG consists of two interpenetrating hyperkagom{\'e} lattices with the triangulated nearest neighbors linked via an exchange interaction, $J_{1}$, while individual triangles are coupled through 2nd nearest neighbor interactions, $J_{2}$ and the two hyperkagom{\'e} lattices linked via a third nearest neighbor term $J_{3}$ \cite{Kinney1979}. In GGG the magnetic Gd$^{3+}$ spins are isotropic ($S$ = 7/2) and are often considered as Heisenberg spins due to single ion anisotropy of less than 0.04~K \cite{Overmeyer}. However the non-negligible dipole exchange, $D$ = 0.7~K,  could lead to anisotropy \cite{Kinney1979}.  Magnetically GGG shows a Curie-Weiss temperature of -2.3~K indicative of AF correlations but does not order down to 0.025~K \cite{Onn1966, Kinney1979}.   Although indications of short range order was hinted at by bulk measurements \cite{Onn1966, Schiffer1995}, proof was obtained by neutron diffraction that revealed  a spin liquid-like ground state down to 0.14~K with the development of  sharper but not resolution limited magnetic diffraction peaks below 0.14~K  \cite{Petrenko1998} in addition to the spin-liquid like scattering. Interestingly long range magnetic order is achieved via the application of only 1~Tesla \cite{Schiffer1994,PetrenkoHFM}. 

The dynamic nature of GGG has previously been studied first via the indirect measurement of $\mu$SR. Two  $\mu$SR studies confirmed the absence of long range order down to 0.025~K, however these studies disagree on the  nature of the slowing down of the spin fluctuations. In a study by Dunsiger {\it et al.} \cite{Dunsiger2000}, a linear decrease of Gd spin fluctuations was observed below 1~K which extrapolated to 8.2~$\mu$eV at 0~K while  Marshall {\it et al.} \cite{Marshall2002}, also determined the slowing down of fluctuations but observed a temperature independent relaxation below 0.2~K. A more direct study, by M\"{o}ssbauer spectroscopy \cite{Bonville2004}, observed the fluctuating Gd spins down to 0.027~K with a decrease in spin fluctuating rate from 11.9~$\mu$eV at 0.4~K to 0.12~$\mu$eV at 0.09~K. Most recently, Ghosh {\it et al.} \cite{Ghosh2008} pointed to a new dynamical phenomena, in the low temperature phase below 0.14~K, in which  fluctuating uncompensated moments coexist with unsaturated AF order and quantum protectorates of defect centered clusters. 

Theoretically Yavorks'kii {\it et al.} \cite{YavorskiiPRL2006} were able to reproduce the spatial correlations of the low temperature (T$<$ 0.14~K) phase by taking into account the nearest neighbor and the nearly infinite dipole exchange interactions, $J$ = 0.107~K \cite{Kinney1979} and $D$ = 0.7 K \cite{PetrenkoPRB_2000}.  Yavorks''kii {\it et al.} showed that $J$ and $D$ are perturbed by  much smaller exchange interactions $J_{2}$ and $J_{3}$  \cite{YavorskiiPRL2006, PetrenkoPRB_2000} and these smaller components dictate the incommensurate ordering wavevector of the low temperature phase T$<$ 0.14~K.
The work in this Letter presents inelastic neutron measurements on polycrystalline GGG from which we obtain both spatial and temporal information \cite{Squires}.  Neutron time-of-flight measurements were performed at the spectrometer IN5 of the  Institut Laue-Langevin \cite{IN5}.   IN5 was set up to an incident energy of $E_{i}$ = 1.94 and 3.27 meV with average elastic linewidths of 50 and 80~$\mu$eV, respectively, full width at half maximum (FWHM). The resolution was determined using a standard incoherent scatterer. The temperature dependence of the scattering function $S( Q, \omega)$ was measured  between 0.06 and 9~K. The instrumental background was measured using an identical empty cell at 2~K and subtracted from the raw data. The sample used in this work is that used in the previous work of Petrenko {\it et al.} \cite{Petrenko1998} containing  99.98 $\%$ of the  non-absorbing isotope $^{160}$Gd. High resolution neutron diffraction using D1A of the Institut Laue Langevin, $\lambda$ = 1.9 \AA{}, was used to determine the upper level of a possible disorder on the Ga/Gd sites. The refinement revealed a fully stochiometric sample, the error of the site occupations indicated the upper limit of disorder to less than 2$\%$. \\
\begin{figure}[htp]
\begin{center}
	\includegraphics[width=10cm, angle=0.]{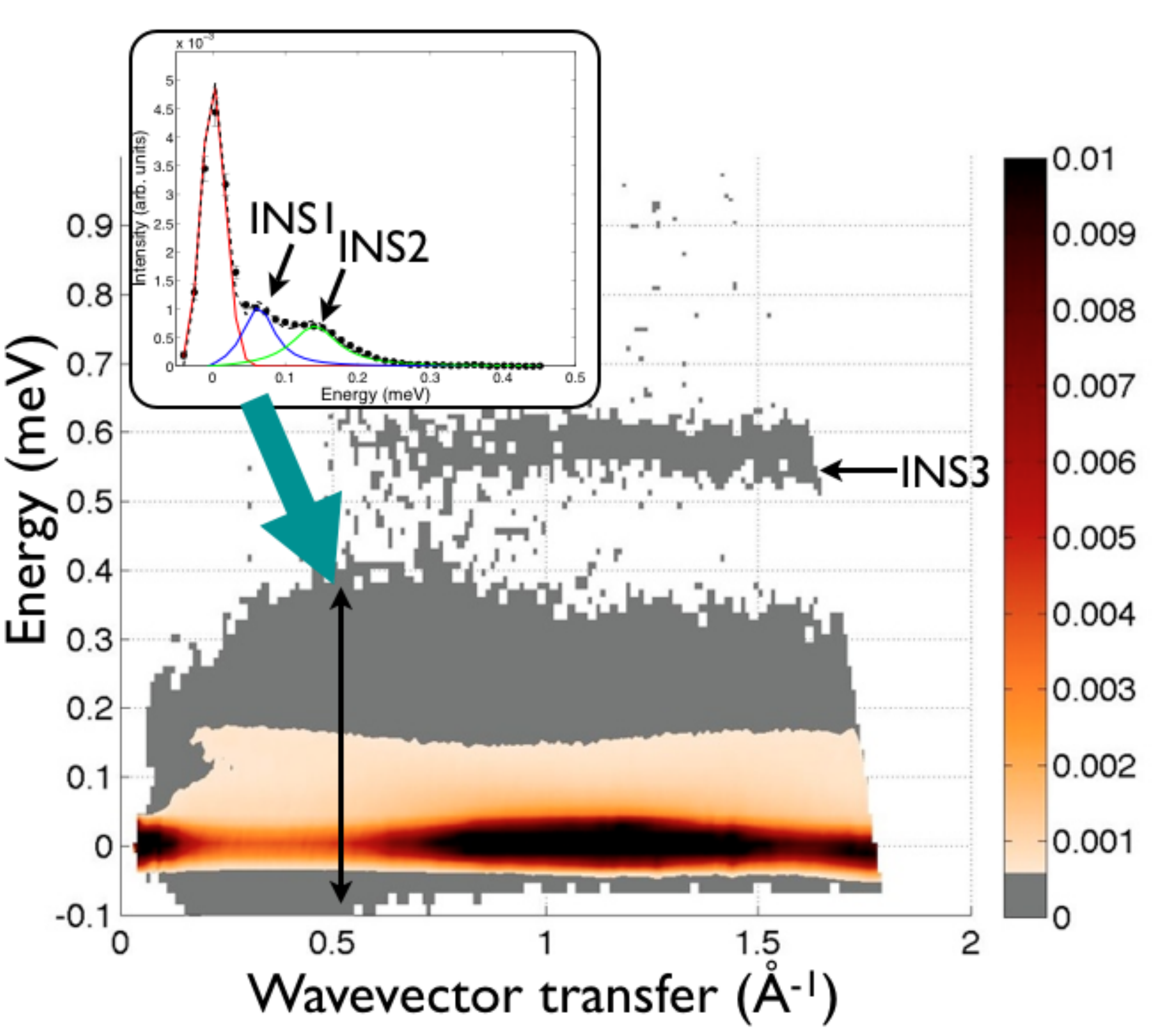}
\caption{Powder averaged scattering function S(Q, $\omega$) of GGG at 0.06~K with incident $E_{i}$ = 1.94~meV. Insert is a cut at wavevector transfer = 0.5~$\AA^{-1}$ showing the elastic lineshape and two low lying excitations with the corresponding fits as described in the text.}
\label{SQWa}
\end{center}
\end{figure}
S( Q, $\omega$) at 0.06~K is shown in  Fig.\ref{SQWa}. Clearly there is a first inelastic contribution at $E$ $=$ 0.58(3)~meV. The lower energy contributions are more easily understood by taking a cut at  constant wavevector transfer, see inset of Fig.\ref{SQWa}, and are well described by a Gaussian elastic line, the FWHM of which is fixed by the vanadium standard, and two further inelastic contributions at 0.04(1) and 0.14(2)~meV. To avoid confusion the three excited states are henceforth named INS1 (0.04~meV), INS2 (0.14~meV) and INS3 (0.58~meV). The inelastic scattering cross sections can be characterized by the dispersion relation, lifetime ($\tau \sim \Gamma^{-1}$) and intensity. These parameters can be obtained by linking the  neutron inelastic magnetic excitation to the dynamic susceptibility via
\begin{equation}
S(Q, \omega) = \frac{1}{\pi} \lbrace n(\omega) + 1) \rbrace F^{2}(Q) \chi^{''} ({\bf Q},  \omega),
\end{equation}
where $F(Q)$ is a dimensionless structure factor that follows the magnetic form factor and  $\lbrace n(\omega) + 1)\rbrace$ is the thermal population factor. The dynamical susceptibility can be further described by a Lorentzian form corresponding to an exponential decay of excitations in time, written in terms of a damped harmonic oscillator:
\begin{equation}
\chi^{''} ( Q,  \omega) =  \frac{4 \omega \omega_{q} \Gamma_{q}}{(\omega^{2} -\Omega_{q}^{2})^{2} +4\omega_{q}^{2} \Gamma_{q}^{2}},
\label{DHO}
\end{equation}
where $\Omega_{q}^{2} = \omega_{q}^{2}+ \Gamma_{q}^{2}$, $\Gamma_{q}$ is a q-dependent linewidth, corresponding to the FWHM of the peak. Furthermore the fit function is convolved with the instrumental resolution \cite{IN5}.
\begin{figure}[htp]
\begin{center}
	\includegraphics[width=10cm, angle=-90.]{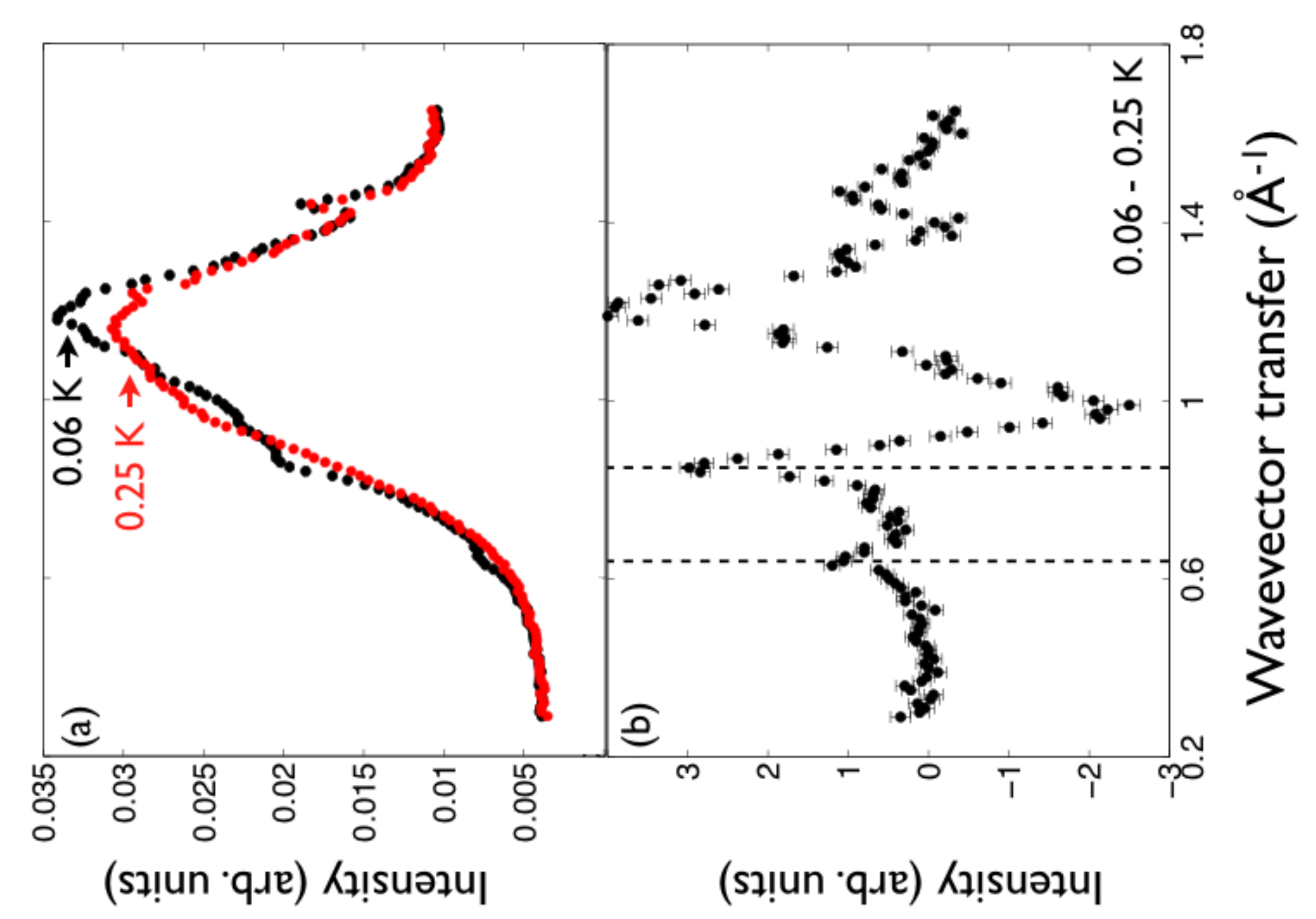}
	\caption{(a) Elastic scattering cross section at 0.06 and 0.025~K. (b) Difference in elastic scattering cross section between 0.06 and 0.25~K. The dashed line corresponds to longish ranged ordered peaks observed by Petrenko {\it et al.} \cite{Petrenko1998}}
\label{Static}
\end{center}
\end{figure}

The data has been characterized at each position of wavevector transfer. Fig. \ref{Static}(a) shows the elastic scattering observed at 0.06 and 0.25~K which represents 82~$\%$ of the total scattering. This scattering is  reminiscent of a spin liquid-like structure factor. In addition,  incommensurate Bragg peaks corresponding to longer range correlations develop below 0.14~K within the elastic line. The resolution of the elastic line gives an upper limit of 50~$\mu$eV for the spin fluctuation rate probed.  Previous $\mu$SR \cite{Dunsiger2000, Marshall2002} and  M\"{o}ssbauer measurements \cite{Bonnville2004} indicate fluctuation rates around 8.2~$\mu$eV which this data will not be sensitive to.  However, the extra scattering that develops below 0.25~K , Fig.\ref{Static} (b),  shows  features of interest, 
marked by the dashed lines, representing the scattering from longer range static correlations first observed by Petrenko {\it et al.} \cite{Petrenko1998} and theoretically reproduced by  Yavorks'kii {\it et al.} \cite{YavorskiiPRL2006}.  Although longer range correlations exist there is no sign of associated spin waves. It is possible that these are too weak to be observed since the correlations remain finite on the scale of 100 \AA{} \cite{Petrenko1998}.  The three inelastic peaks, INS1, INS2 and INS3 are all dispersionless within the resolution probed. These excitations do not originate from either local vibrational excitations nor crystal field excitations since their dependence on wavevector transfer neither increases with $|Q|$, as would be expected from local vibrational excitations, nor follow the Gd$^{3+}$ form factor expected for crystal field excitations \cite{BrownTables, Squires}. The possibility that the higher level excitation is a crystal field excitation affected by an internal molecular field can be excluded as this is not compatible with the specific heat data \cite{Kinney1979, BrianPrivate}. 

\begin{figure}[htp]
\begin{center}
	\includegraphics[width=7cm, angle=0.0]{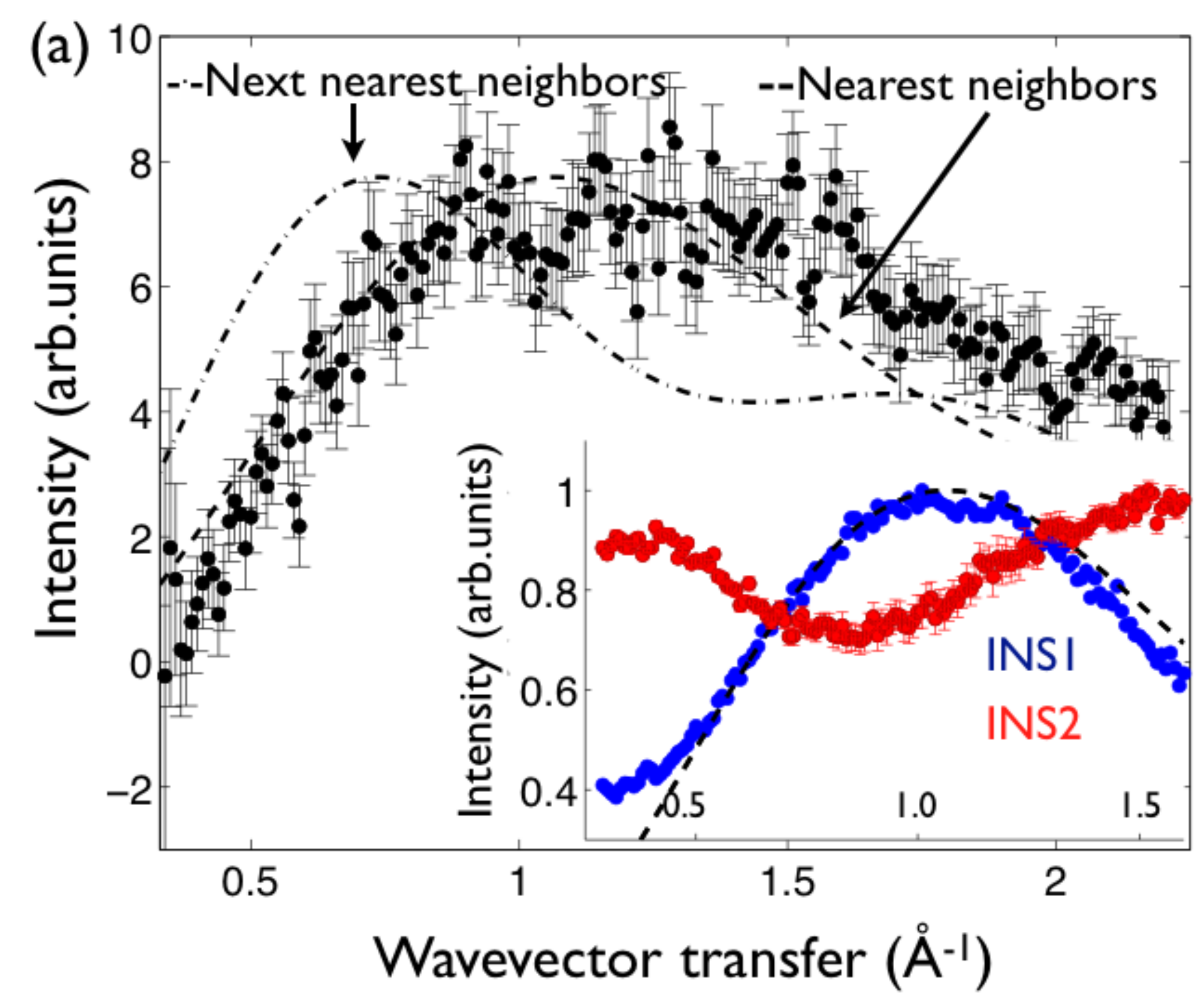}
	\includegraphics[width=7cm, angle=0.0]{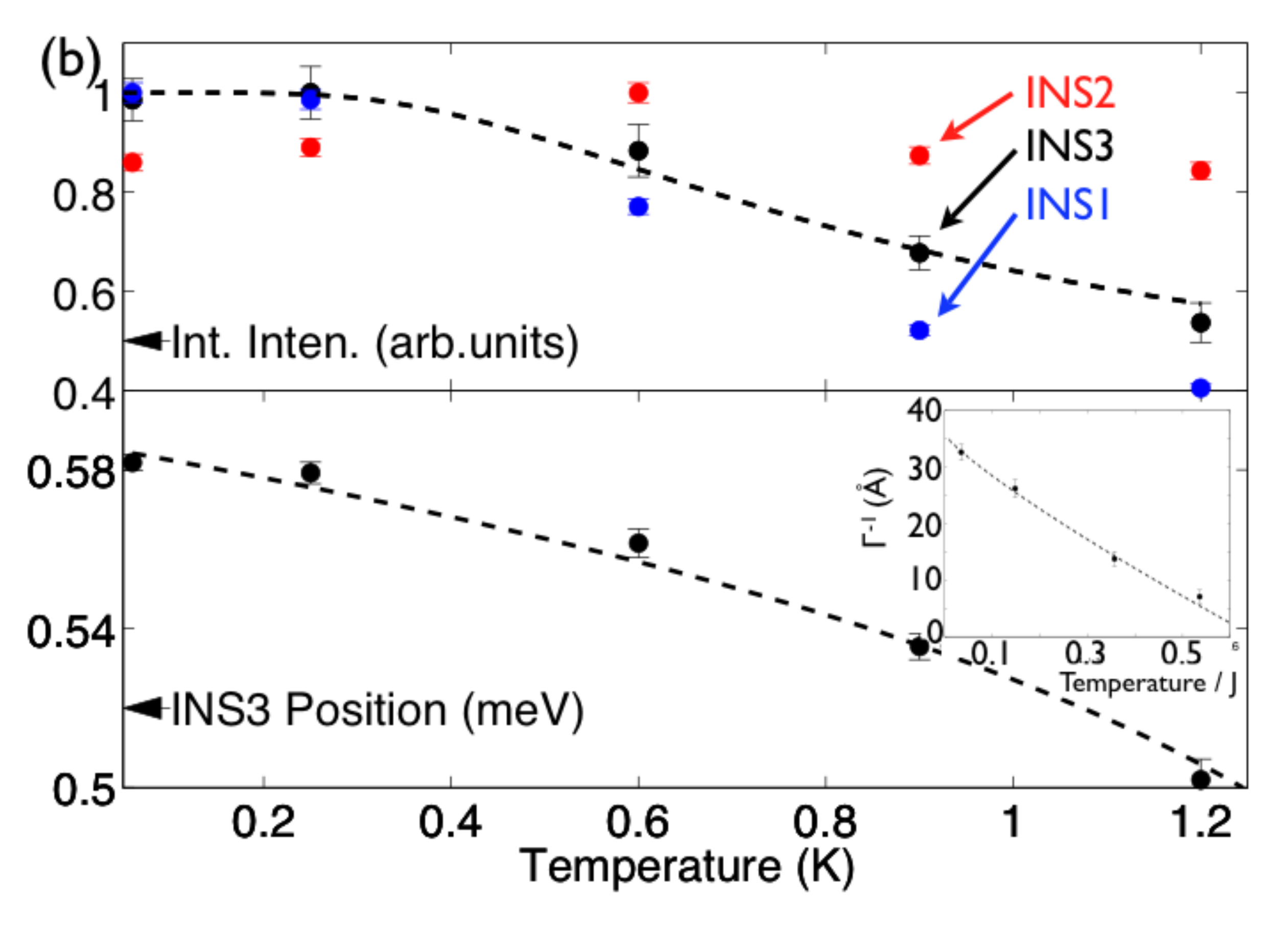}
	\caption{(a) Integrated intensity of INS3 as a function of wavevector transfer. The dashed lines correspond to a model of short range AF correlation with only near neighbor (--) or next nearest neighbor correlations (-$\cdot$-). The inset figure shows the integrated intensities of the two low lying excitations INS1 and INS2.  (b)(top)Temperature dependence of the integrated intensity for INS1, INS2 and INS3.  The dashed line is the temperature dependence expected for a singlet to triplet excitation with an exchange interaction $J_{GdNN} = J_{NN}S(S+1) = 1.68 K$ and follows closely the integrated intensity of INS3.  (b)(bottom) Peak position in energy of INS3 (dashed line is a fit to the data with a power law corresponding to a transition temperature, T = 1.67(2) K.), inset: T/J dependence of excitation lifetime of INS3. Dashed line is the algebraic function $\tau =  \mathcal{A} T^{\xi}$ with an exponent $\xi$ = 0.84 $\pm$ 0.21}
\label{Dimer}
\end{center}
\end{figure}

Fig.\ref{Dimer}(a) shows the wavevector dependence of the integrated intensity of the three inelastic peaks. The intensity of INS3 at 0.06~K, integrated in energy across  the region of interest, reveals spatial correlations corresponding to the scattering cross section expected from a  triplet excitation above a ground state of singlet dimers \cite{Furrer1977}.  In this case, a dimer can be understood as short range order of AF coupled spins within a cluster effectively shielded from its neighboring cluster. Magnetic interactions between clusters can therefore be neglected. The neutron scattering cross section for such a ground state is given by 
\begin{equation}
\frac{d^{2} \sigma}{d \Omega d \omega} \propto A(T) F^{2}(Q)\left[1 - \frac{\sin(Qd)}{Qd}\right], 
\label{Eqnimer}
\end{equation}
where $d$ is the separation between spins,  $F(Q)$ is the Gd$^{3+}$ magnetic form factor \cite{BrownTables} and $A(T)$ is  a temperature scaling factor linked to the canonical partition function proportional to the thermal distribution of the singlet ground state and the triplet excited state, $A(T) = 1/(1+3$exp$(-J_{GdNN}/k_{B}T))$, with $T$ = temperature, $k_{B}$ the Boltzmann constant and $J_{GdNN}$ is the nearest neighbor exchange energy $J_{GdNN} = J_{NN}S(S+1) = 1.68 K$ \cite{Furrer1977}. The dashed lines in Fig.\ref{Dimer}(a) correspond to spatial correlations with near neighbor exchange interaction (- -), $d$ = 3.7915 ~\AA{} and next nearest exchange interaction (-$\cdot$-) $d$ = 5.7916~\AA{}. Clearly the data are well described by a model including only near neighbor exchange interactions.  The energy FWHM of this excited state at 60 mK is equivalent to the instrumental resolution, thus indicating strong coupling between dimerized Gd moments. The inset of Fig. \ref{Dimer}(a) shows the integrated intensity of the wavevector transfer of INS1, INS2, obtained with incident neutron energy $E_{i}$ = 1.94~meV to optimize resolution. The variation of width and position of the peaks do not exceed the resolution of the instrument and are thus considered as constants. The wavevector dependence of the integrated intensity of INS1 follows closely the short range order  behavior of INS3, the dashed line represents the lineshape of Eqn. \ref{EqnDimer}.  At low and high wavevector transfer the model of short range correlations fails indicating that extra terms remain important for a full description of the ground state. The integrated intensity of INS2 shows a minima close to the position in reciprocal space that corresponds to nearest  neighbor correlations indicating that the origin of INS2 is very different to that of INS1 and INS3. 

The temperature dependence of the normalized integrated intensities of INS1, INS2 and  INS3 is shown in Fig.\ref{Dimer}(b)(top). The dashed line represents the thermal behavior expected from a singlet-triplet excitation, $A(T)$, with an exchange interaction $J_{GdNN} =$~1.68~K, the near neighbor exchange interaction obtained by Schiffer {\it et al.}  \cite{Schiffer1995}. The dashed line follows closely the integrated intensity of INS3 thus further validating the notion of a dimerized short range AF ordered state.  Neither INS1 nor INS2 follow the temperature dependence of INS3. The integrated intensity of the INS1 excitation follows a trend similar to INS3, albeit with a reduced exchange interaction J = 1.3 K,  up to 0.6 K but at higher temperatures does not follow this trend. The integrated intensity of INS2 has a maxima at 0.6 K. 

Further information concerning the INS3 excitation is revealed in Fig. \ref{Dimer}(b)(bottom). Unlike INS1 and INS2, INS3 shows a strong temperature dependence in its energy position and the excitation lifetime.  The peak energy position can be followed by a  power law function with parameters $\beta = 0.12(1)$ and falls to zero at  T = 1.67(2)~K.   The excitation lifetime can be described by the algebraic form $\tau =  \mathcal{A}$ T$^{\xi}$ with $\xi$ = 0.84 $\pm$ 0.21 and $\mathcal{A}$ diverges to the inverse of the instrument resolution at 0.06~K, see the inset of  Fig. \ref{Dimer}(b)(bottom). The relevance of these parameters becomes clear when reviewing recent theoretical work by Robert {\it et al.} \cite{Robert2008} showing that, in contrast to the pyrochlore lattice \cite{Moessner1998, Moessner1998PRL, Conlon2009}, sufficient temporal and spatial stiffness  in a classical kagom{\'e} antiferromagnet can give rise to magnetic excitations  corresponding to acoustic and optical modes in addition to a soft mode.

These excitations depend strongly on the temperature regime.  At high temperatures ($T/J$ $\geq$ 0.2 ) only a quasielastic signal centered at 0 meV is expected.  On decreasing $T/J$ from 0.2 to $10^{-2}$ the quasielastic signal splits into two excitations, an acoustic mode and a non-dispersive soft mode that softens with decreasing temperature  below $T/J$ $\leq$ $10^{-2} $.  
An optical mode develops at $\omega$ = $2J$ for $T/J$ $<$ $5$e$-3$. Furthermore, the theoretical excitations are characterized by the temperature dependence of their lifetime with an algebraic dependence $\tau = \mathcal{A} T^{\xi}$ with $\xi$ = 0.995 for quasielastic scattering and $\xi$ = 0.18 for inelastic scattering in the regime of cooperative paramagnetism, $T/J$ $\leq$ 0.1.  The temperature regime probed in this work extends from the partially ordered phase at 0.06~K into the paramagnetic regime at 1.2 K corresponding to 0.04$<$$T/J$$<$0.7 (with $J$ = 1.68 ~K). 

The excitations observed cannot be assigned to acoustic excitations which would remain dispersive and originate from Bragg peaks, even with powder averaging. However it is possible to assign the non-dispersive gapped excitations to the optical or soft modes with  a high energy mode at $\omega$ = $4J$ and not $2J$ as predicted \cite{Robert2008}.  Robert calculated the excitations for a classical Heisenberg kagom{\'e} AF. It is well known that substantial long range dipole exchange interactions play an important role, such a mode would therefore be lifted upwards due to further exchange interactions and could therefore be observed at  $\omega$ = $4J$.  Analogous to phonons, a magnetic optical mode can arise from a localized perturbation of interactions as found in the short range dimerized interactions displayed by INS1 and INS3.

The data presented in this work sheds light on the unusual magnetic ground state of the hyperkagom\'{e} structure GGG. The data indicate that the longer range order, observed below 0.14 K in previous diffraction work is static and does not impact on the behavior of the higher energy spectral density which can be linked to the partially ordered state of GGG. 82 $\%$ of the scattering is static with a fluctuation rate of less than 0.05~$\mu$eV incorporating a large component of the spin liquid-like structure factor. The remaining spectral weight lies in three gapped magnetic excitations two of which can be modeled with the spatial dependence of short range AF dimer-like correlations.  This is highly unusual in a compound with a ground state manifold in which  a continuum of excitations  is expected to be characteristic of the dynamic nature of the magnetic ground state, thus leading to a new class of magnetic dynamic order for hyperkagom\'{e} compounds.

\bibliography{Gd3Ga5O12_PRL}

\end{document}